\documentstyle[aps,prb,twocolumn,epsfig]{revtex}
\begin{document}
\title{Left-handed materials}
\author{P. Marko\v{s}$^{1,2}$ and C. M. Soukoulis$^{1,3}$}
\address{$^1$ Ames Laboratory and Department of Physics and Astronomy,
Iowa State University, Ames, Iowa 50011\\
$^2$ Institute of Physics, Slovak Academy of Sciences,
   D\'ubravsk\'a cesta 9, 842 28 Bratislava, Slovakia\\
$^3$ Research Center of Crete, 71110 Heraklion, Crete, Greece}
\maketitle

\def\eps{\epsilon}
\def\em{\epsilon_{\rm m}}
\def\ed{\epsilon_{\rm d}}
\def\epseff{\epsilon_{\rm eff}}
\def\mueff{\mu_{\rm eff}}
\def\epse{\epseff}
\def\mue{\mueff}
\def\ne{n_{\rm eff}}

\def\be{\begin{equation}}
\def\ee{\end{equation}}

\section{Introduction}\label{ms-1}
Rapidly increasing interest in the left-handed materials
(LHM) started after   Pendry
{\it et. al.}  predicted that certain
man-made composite structure could  possess,  in a given frequency interval, a
negative {\it effective} magnetic permeability $\mueff$ \cite{Pendry-1}.
Combination of such a  structure with negative effective
permittivity medium - for instance the
regular array of thin metallic wires
\cite{Pendry-2a,Pendry-JPCM,Sigalas,Soukoulis,Sarychev,Efros-1} - enabled
the construction  of meta-materials
with both {\it effective} permittivity and permittivity {\it negative}.
This was confirmed by  experiments \cite{Smith,Shelby-APL}.

Structures with negative permittivity and permittivity were named 
``left-handed''
   by Veselago \cite{Veselago}
over 30 years ago to emphasize  the fact that the intensity of the 
electric field $\vec{E}$,
the magnetic intensity $\vec{H}$ and the wave
vector $\vec{k}$ are related by a left-handed rule. \cite{pozn}
This can be easily seen by writing Maxwell's equation for a plane 
monochromatic wave:
$\vec{k}\times\vec{E} = \frac{\omega\mu}{c} \vec{H}$ and
$\vec{k}\times\vec{H} = -\frac{\omega\epsilon}{c}\vec{E}$
Once $\epsilon$ and $\mu$ are both positive, then $\vec{E}$, 
$\vec{H}$ and $\vec{k}$
form a right set of vectors.  In the case of negative $\epsilon$ and 
$\mu$, however, these three
vectors form a left set of vectors.

In his pioneering work, Veselago  described the physical properties 
of LH systems:
Firstly, the direction of the energy flow, which is given by the 
Poynting vector
\begin{equation}\label{poynting}
\vec{S} = \frac{c}{4\pi}~\vec{E}\times\vec{H}
\end{equation}
does not depend on the sign of the permittivity and permeability
of the medium. Then,
the vectors $\vec{S}$ and $\vec{k}$  are parallel (anti-parallel) in
the  right-handed (left-handed) medium, respectively.
Consequently, the phase and group
velocity of an electromagnetic wave propagate in {\it opposite} 
directions  in the
left-handed material. This
gives  rise to a number of novel physical phenomena, as were discussed already
by Veselago. For instance, the Doppler effect and  the Cherenkov
effect are reversed in the LHM \cite{Veselago}.

If both $\eps$ and $\mu$ are negative, then also the refraction index 
$n$ is negative
\cite{Veselago,SK}.
This  means the negative refraction of the electro magnetic wave
passing  through the boundary
of two materials, one with positive and the second with negative $n$
  (negative Snell's law).
Observation of negative Snell's law,  reported experimentally
  \cite{Shelby-Science} and later
in \cite{Claudio-1}, is today a subject of rather controversially debate
\cite{Walser,Pendry-answer,Garcia_1,Sanz}.
Analytical arguments of the sign of the refraction index
were presented in \cite{SK}.
Numerically, negative phase velocity   was observed  in FDTD simulations
\cite{Ziolkowski}.
Negative refraction index was  calculated from the transmission and 
reflection data
\cite{MS-3}. Finally, negative refraction
on the wedge experiment was demonstrated also by FDTD simulations 
\cite{Claudio-private}.

Negative refraction allows the fabrication of flat lens \cite{Veselago}.
Maybe the most challenging property of the left-handed
medium is its ability to enhance the
evanescent modes \cite{Pendry_3}. Therefore  flat lens,
constructed from left-handed material with
$\eps=\mu=-1$  could in principle work as
perfect lens  \cite{Pendry_3} in the sense that it can reconstruct  an
object without any diffraction error.

The existence of the perfect lens
seems to be in contradiction with fundamental physical laws, as was 
discussed in a series of papers
\cite{Walser,comments,Garcia_2}.
Nevertheless, more detailed  physical considerations
\cite{Pendry-answer,reply,Pendry-Garcia,Lu,Gomez} not only showed 
that the construction of ``almost perfect'' lens
is indeed possible, but brought some more insight into this phenomena
\cite{Ruppin,Haldane,SS,RPS,Feise,Nefedov,Zhang}.

As Veselago  also discussed in his pioneering paper,
the  permittivity and the permeability of the left-handed material 
must depend on the
frequency of the EM field,  otherwise the energy density \cite{Landau}
\be\label{11}
U=\frac{1}{2\pi}\int
d\omega\left[\frac{\partial(\omega\eps')}{\partial\omega}|E|^2
+\frac{\partial(\omega\mu')}{\partial\omega}|H|^2\right]
\ee
would be {\it negative} for {\it negative}
$\eps'$ and $\mu'$
(real part of the permittivity and permeability).
Then, according to
Kramers- Kronig relations,  the imaginary part of the permittivity ($\eps''$)
and of the permeability ($\mu''$) are non-zero in the LH materials.
Transmission losses are therefore unavoidable
in any LH structure. Theoretical  estimation of losses is rather 
difficult problem,
and led even to the conclusion that LH materials are not transparent
\cite{Garcia_1}. Fortunately, recent experiments
\cite{Claudio-1,Claudio-2} confirmed the  more optimistic theoretical 
expectation
\cite{MS-4}, that the losses in the LH structures might be
as small as in conventional
RH materials.

The number of papers about left-handed  materials increased
dramatically last year.
The present paper is not
the first review about left-handed materials. For recent reviews, see
\cite{Smith-nato} or the papers of Pendry \cite{Pendry-2}.
Here we present typical  structures of the left-handed  materials 
(Sect. \ref{ms-2}),
discuss a numerical method of simulation
of the propagation of EM waves based on the transfer matrix (Sect. 
\ref{ms-3}), and present some recent results
obtained by this method (Sects. \ref{ms-4}, \ref{ms-5}). We  discuss
in Sect. \ref{ms-5} how the  transmission depends
on various structural and material parameters of LH  structure.
The method of calculation of the
refractive index and of the effective permittivity and permeability 
is presented  and applied to the
LH structure. An unambiguous proof of the negative refraction index 
is given and the {\it effective}
permittivity and permeability are calculated in Sects. \ref{ms-6}, 
and \ref{ms-7}.
The obtained data for the permittivity  are rather
counter-intuitive and require  some  physical interpretation. Finally, in the
Section \ref{ms-8} we discuss some new directions of the development of
both theory and experiments.

\begin{figure}[t!]
\begin{center}
\epsfig{file=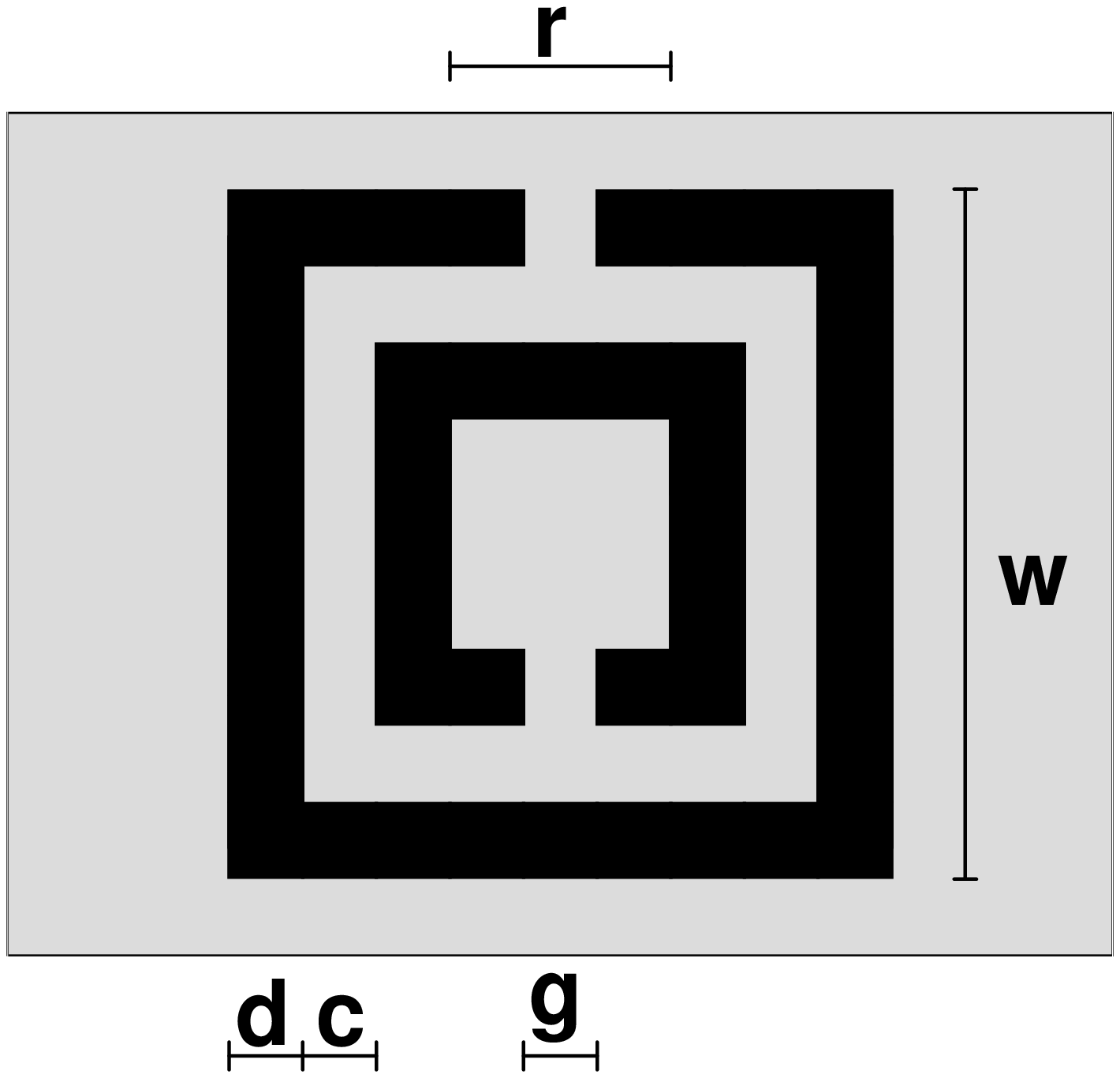,width=0.18\textwidth}
\epsfig{file=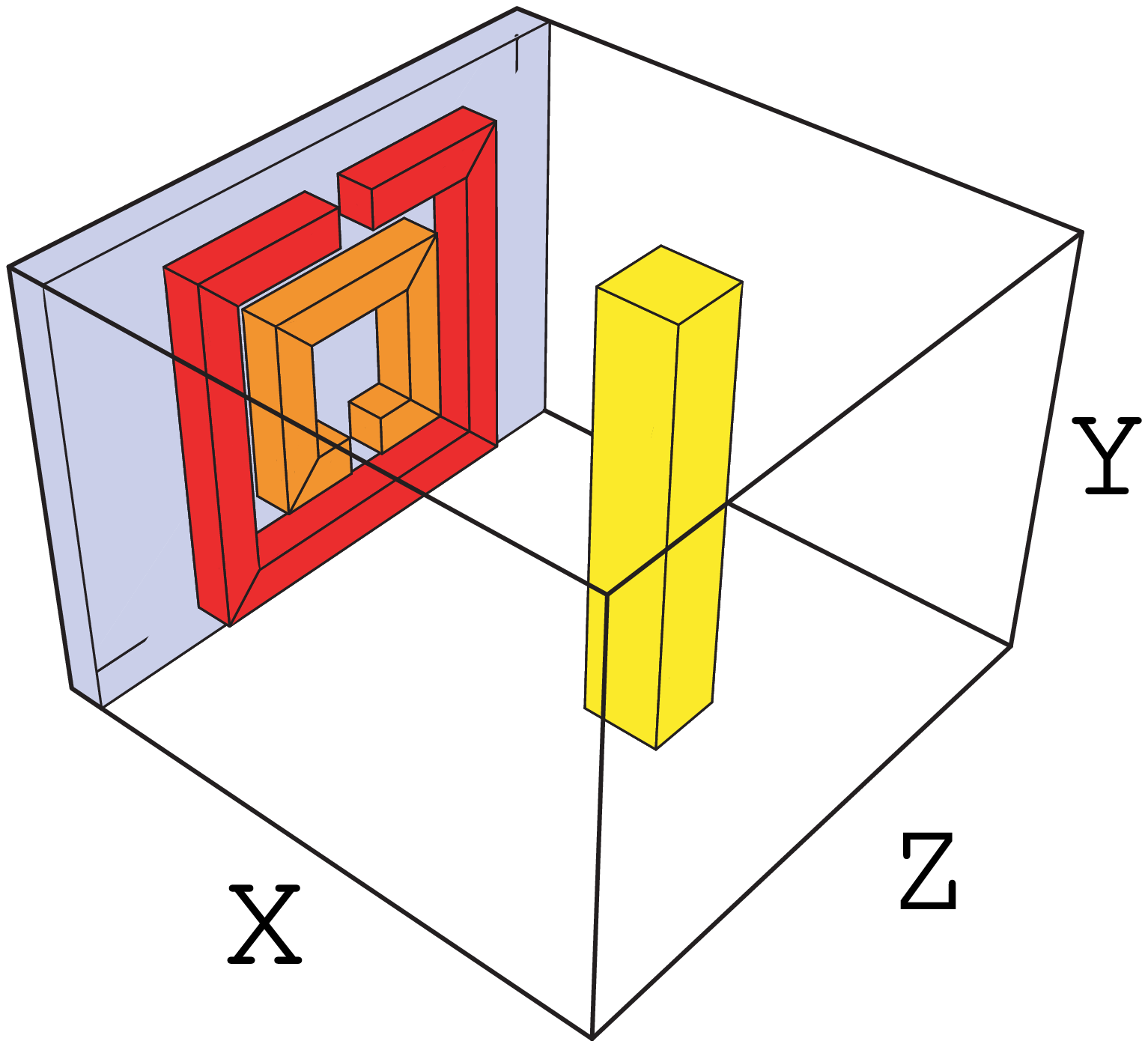,width=0.25\textwidth}
\end{center}
\caption{Left: Structure of the split ring resonator (SRR).
The SRR consists of two splitted metallic
``rings''. The SRR is characterized by the size $w$,
width of the rings $d$, and two gaps: $g$ and $c$. The external 
magnetic field induces an
electric current in both rings \cite{Pendry-1}.  The
shape of the SRR  (square or circular)
is not crucial for the existence of the magnetic resonance.
Right: Structure of the unit cell of the left-handed material.
Each unit cell contains the split ring resonator located on the
dielectric board, and one wire.  Left-handed structure is created by regular
lattices of unit cells.
The EM wave propagates along the $z$ direction.
Periodic boundary conditions are considered in the $x$ and
the $y$ direction, which assures the periodic distribution of the EM field.
}
\label{ms-f1}
\end{figure}

\section{Structure}\label{ms-2}

LHM materials are by definition composites, whose
properties are not determined by the fundamental physical  properties
of their constituents but by the
shape and distribution of specific patterns included in them.
The route of the construction of  the of LH structure consists from 
three steps:

Firstly, the   split ring resonators (SRR) (see fig. \ref{ms-f1} for 
the structure of SRR)
was predicted to exhibit the resonant
{\it magnetic} response to the EM wave, polarized with $\vec{H}$ 
parallel to the axis of the SRR.
Then, the periodic array of SRR is characterized \cite{Pendry-1} by
the {\it effective} magnetic permeability
\begin{equation}\label{mueff}
\mueff(f)=1- \frac{F \nu^2}{\nu^2-\nu_{m}^2+i\nu\gamma}~.
\end{equation}
In (\ref{mueff}), $\nu_{m}$ is  the resonance frequency which depends 
on the structure of the SRR
(fig. \ref{ms-f1}) as
$(2\pi \nu_{m})^2=3L_xc_{\rm light}^2/[\pi\ln(2c/d)r^3]$.
$F$ is the filling factor of the SRR within one unit cell
and  $\gamma$ is
the damping factor $2\pi\gamma=2L_x\rho/r$, where $\rho$ is the 
resistivity of the metal.

Formula (\ref{mueff}) assures that the   {\it real} part of $\mueff$ 
is {\it negative}
at an interval $\Delta\nu$ around the
resonance frequency.
If an array of SRR is combined with a medium with negative
{\it real} part of the permittivity, the resulting structure would possess
{\it negative effective} refraction index in the resonance frequency 
interval $\Delta\nu$
\cite{SK}.

The best candidate for the negative permittivity medium
is a regular lattice of thin metallic wires, which  acts as a high pass
filter for the EM wave polarized with $\vec{E}$ parallel to the 
wires. Such an array
exhibits  negative  effective permittivity
\begin{equation}\label{epseff}
\epseff(\nu)=1-\frac{\nu_p^2}{\nu^2+i\nu\gamma}.
\end{equation}
\cite{Pendry-2a,Sigalas,Sarychev}
with the plasma frequency
$\nu_p^2=c_{\rm light}^2/(2\pi a^2\ln(a/r))$
\cite{Pendry-2a}. Sarychev and Shalaev derived another expression for 
the plasma frequency,
$\nu_p=c^2_{\rm light}/(\pi a^2{\cal L})$  with ${\cal 
L}=2\ln(a/\sqrt{2}r)$$+\pi/2-3$ \cite{Sarychev}.
Apart from tiny differences in both formulas, the two theories are 
equivalent \cite{Efros-1}
and predict that effective permittivity is {\it negative} for $\nu<\nu_p$.

By combining  both the above structures,
a left-handed  structure can be created. This was done for the first 
time in the experiments of
Smith {\it et al.} \cite{Smith}. Left-handed material  is a periodic structure.
A typical unit cell of the left-handed structure is shown in fig. \ref{ms-f1}.
Each unit  cell contains a metallic wire and one split ring resonator
(SRR), deposited on the dielectric board.

\begin{figure}[t!]
\begin{center}
\epsfig{file=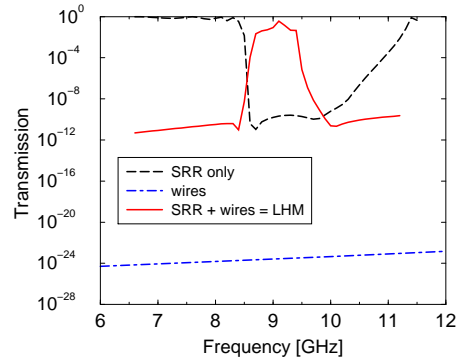,width=0.33\textwidth}
\end{center}
\caption{Transmission of the EM wave, polarized with 
$\vec{E}\!\parallel\! y$ and $\vec{H}\!\parallel\! x$,
through a periodic array of split ring resonators, wires, and of both 
SRR and wires.}
\label{ms-f2}
\end{figure}

\medskip

Fig. \ref{ms-f2}  shows the transmission of the EM waves through the 
left-handed structure
discussed above.
The transmission through the array of the SRR is close to unity for 
all the frequencies outside the
resonance interval (8.5-11 GHz in this particular case) and decreases
to -120 dB in this interval, because   $\mueff$ is negative
(Eq. \ref{mueff}).
The transmission of the array of metallic wires is very small for all 
frequencies
below the plasma frequency (which is $\sim 20$ GHz in this case), 
because $\epseff$ is negative
(Eq. \ref{epseff}). The structure
created by the combination of an  array of SRR and wires exhibits 
high transmission $T\sim 1$
within the resonance interval, where both $\epseff$ and $\mueff$ are  negative.
For frequencies outside the resonance interval, the product 
$\mueff\epseff$ is negative.
The transmission decays with the system length, and is only
$\sim -120$ dB in the example of fig. \ref{ms-f2}.
Experimental analysis of the transmission of all the three structures was
performed by Smith {\it et al.}  \cite{Smith}.

We want to obtain a resonance frequency $\nu_m\approx 10$ GHz. This
requires the size of the unit
cell to be 3-5 mm. The wavelength of the EM wave with frequency $\sim 
10$ GHz is $\approx 4$
cm, and exceeds by a factor of 10 the structural details of the 
left-handed  materials. We can
therefore consider
the left-handed  material as macroscopically homogeneous.
This is the main difference between the left-handed structures and
the ``classical'' photonic band gap  (PBG) materials, in which the 
wave length is
comparable with the lattice period.

\begin{figure}[t!]
\begin{center}
\epsfig{file=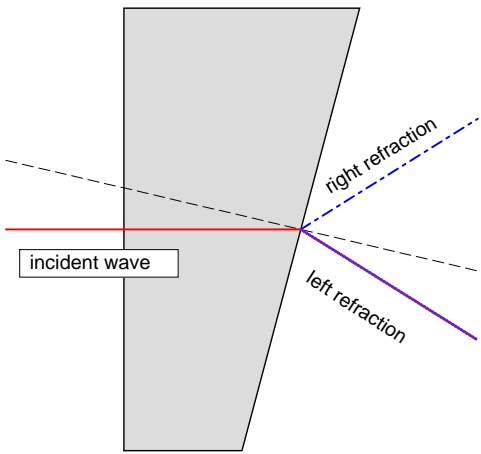,width=0.35\textwidth}
\end{center}
\caption{Refraction experiment on the Left-handed material 
\cite{Shelby-Science,Claudio-2}.
The incident EM wave propagates from the left and hits perpendicularly
the left boundary of the wedge. The angle of refraction is measured
when the EM wave passes the right boundary of the inspected material 
and propagates for some time
in the air.  Two possible directions of the propagation of the
refracted wave are shown: the right refraction for the conventional 
right-handed (RH) material, and
left refraction for the left-handed material. This experimental 
design enables to use also strongly anisotropic
one-dimensional LH samples, since the angle of refraction is measured 
{\it outside} the sample.
}
\label{ms-f3}
\end{figure}

It is important to
note that the structure described in fig. \ref{ms-f1}
is strongly anisotropic.  For frequencies
inside  the resonance interval, the effective
$\epseff$ and $\mueff$ are  negative only for EM field with 
$\vec{H}\!\parallel\! x$ and
$\vec{E}\!\parallel\! y$. The left-handed properties appear only when 
a properly polarized
EM wave propagates in the $z$ direction.
The structure in fig. \ref{ms-f1} is therefore {\it effectively} 
one-dimensional.
Any EM waves, attempting to propagate either along the $x$ or
along the $y$ direction would decay exponentially since the corresponding
product $\epseff\mueff$ is negative.
This structure is therefore not suitable
for the realization of the perfect lens.
To test the negative Snell's law experimentally, a wedge type of experiment
must be considered \cite{Shelby-Science} (fig. \ref{ms-f3})
in which the angle of refraction is measured {\it outside}, the 
left-handed medium  (in air)
\cite{Shelby-Science,Claudio-1}.

Two dimensional structures have also been constructed.  For instance,
the anisotropy in the $x-z$ plane is removed
if each unit cell contains two SRR located in two perpendicular planes
\cite{Shelby-APL,Shelby-Science}.  No three dimensional structures
have been experimentally  prepared yet.

\begin{figure}[t!]
\begin{center}
\epsfig{file=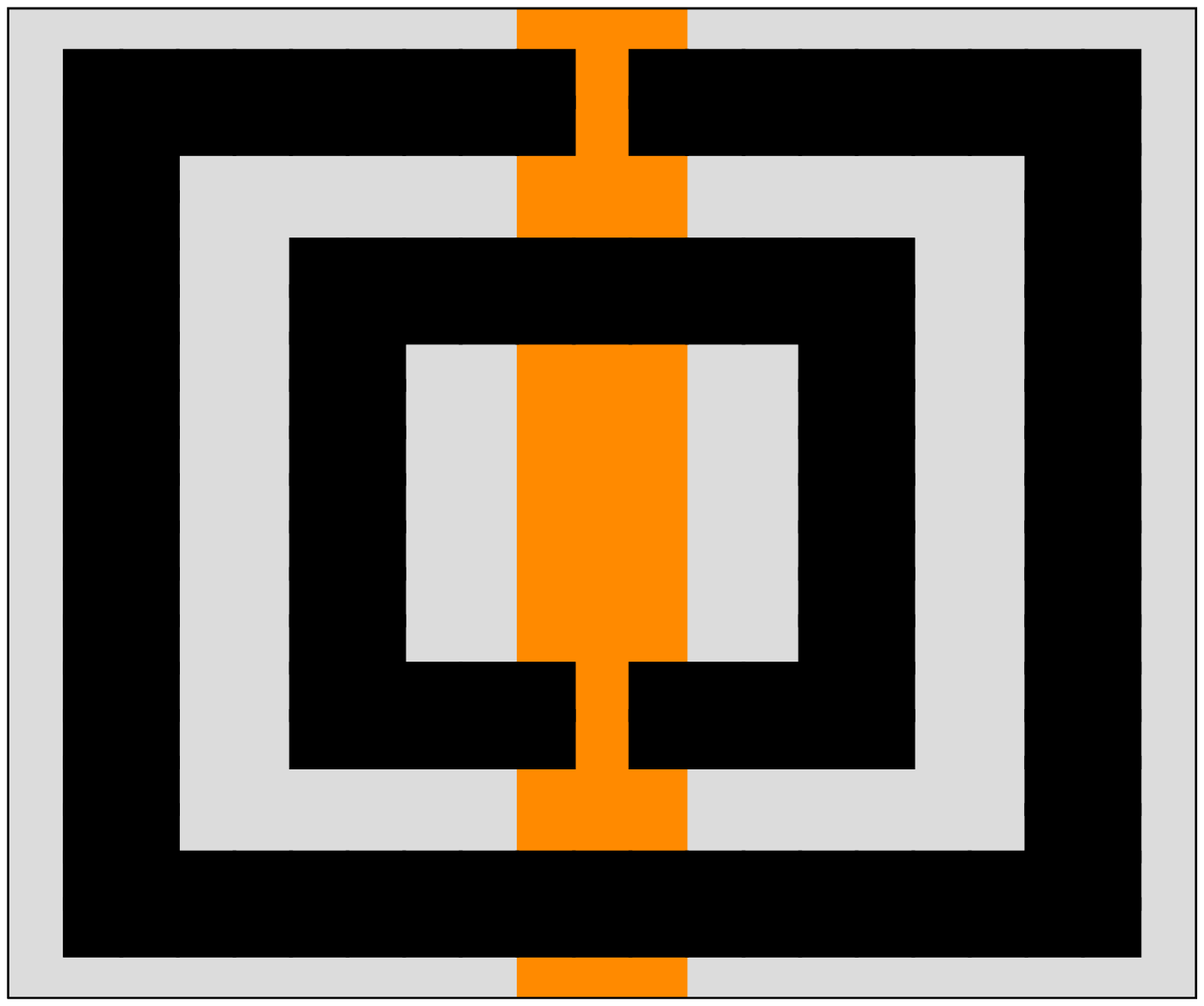,width=0.22\textwidth}
\epsfig{file=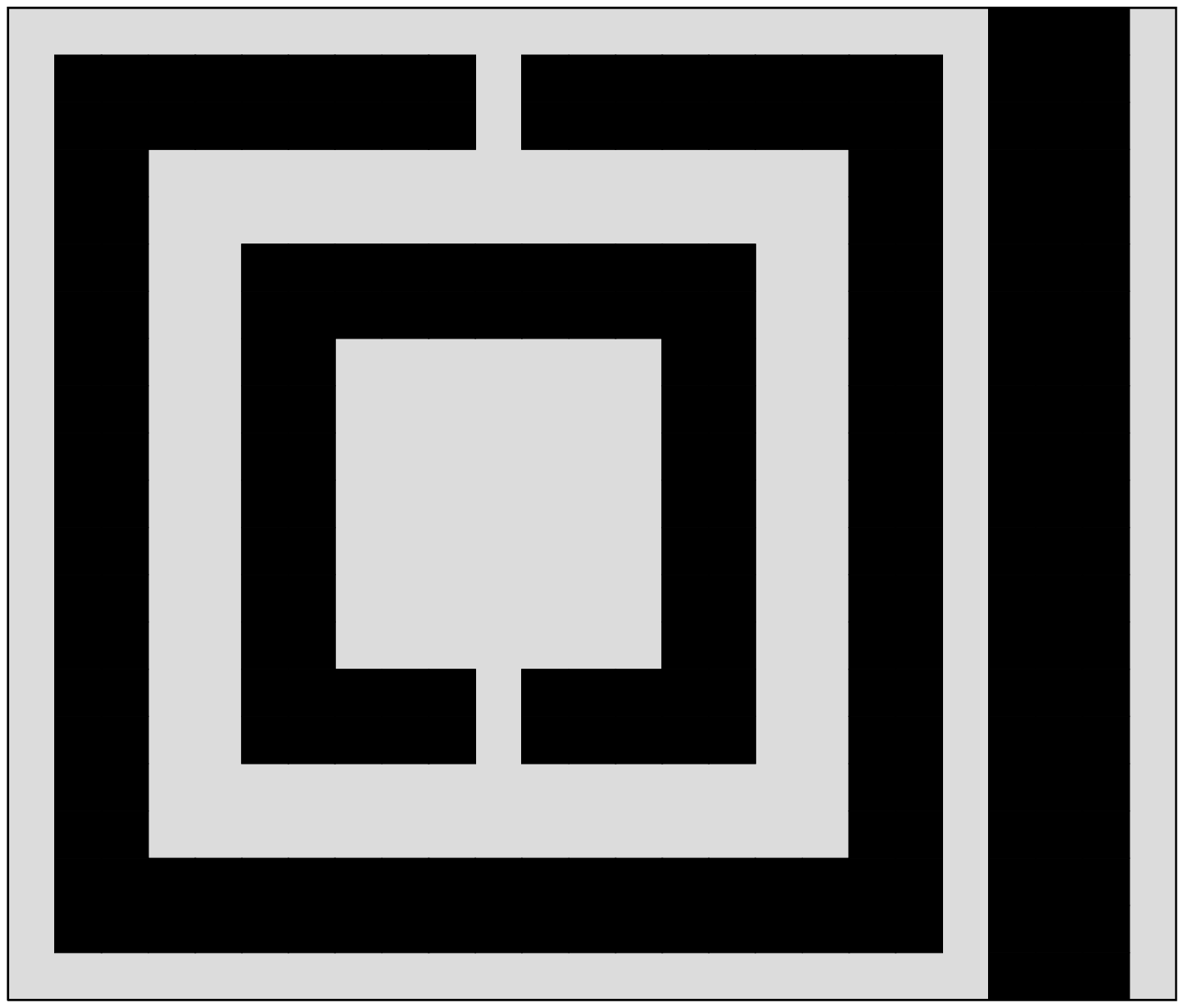,width=0.22\textwidth}
\end{center}
\caption{Two positions of the metallic wire in the unit cell. Left: 
wire is deposited on the
{\it opposite} side of the dielectric board 
\cite{Shelby-APL,Claudio-1}. Right: wire is located
along the split-ring resonator \cite{MS-5}.}
\label{ms-f4}
\end{figure}

It is also worth mentioning that some other one dimensional
structures were prepared,
in which the wires  were deposited on the same side of the dielectric
board with the SRRs.
The wires could be located either on the opposite side of the
dielectric board, as it was done
in \cite{Shelby-APL,Shelby-Science,Claudio-2}, or put next to the SRRs on the
same side of board
\cite{MS-5}  as shown in the right panel of fig. \ref{ms-f4}. More 
complicated one-dimensional
structure was suggested by
Ziolkowski \cite{Ziolkowski-preprint}. Recently, Marques {\it et al.} 
found left-handed
behavior in an  array of SRR located inside a metallic wave guide 
\cite{Marques-2}.

\section{Numerical simulation}\label{ms-3}

Various numerical algorithms were used to simulate the propagation of 
EM waves through the
LH structure. We concentrate on the transfer matrix algorithm, 
developed in a series of papers by Pendry
and co-workers \cite{Pendry-TM}. The transfer matrix algorithm 
enables us to calculate
the transmission, reflection, and absorption as a function of 
frequency \cite{MS-1,MS-2}. Others
use commercial software: either Microwave studio 
\cite{Clausio-1,Weiland,Claudio-2}
or MAFIA \cite{Smith,Smith-nato}, to estimate the position of the
resonance frequency interval. Time-dependent analysis, using various 
forms of FDTD
algorithms are also used \cite{Claudio-1,Ziolkowski,Ziolkowski-preprint,GP,Kik}.

In the transfer matrix algorithm,
we attach in  the $z$ direction, along which EM wave propagates, two 
semi-infinite
ideal leads with $\eps=1$ and $\mu=1$.
The length of the system varies from 1 to 300 unit cells.
Periodic boundary conditions along the $x$ and $y$
directions are used.
This makes the system effectively infinite in the transverse directions, and
enables us to restrict the simulated structure to only one unit cell 
in the transverse
directions.

A typical size of the unit cell is 3.66 mm.
Because of numerical problems,
we are not able to treat very thin metallic structures. While in
experiments the thickness of the SRRs is
usually  17 $\mu$m, the thickness used in the numerical calculations
is determined by the
minimal mesh discretization, which is  usually $\approx 0.33$ mm.
In spite of this constrain, the
numerical data are in qualitative  agreement with the experimental
results. This indicates that the
thickness of the SRR is not a crutial parameter, unless it decreases
below or is comparable with the skin
depth $\delta$. As $\delta\approx 0.7 \mu$m at GHz frequencies of
interest,  we are far from this
limitation. We will discuss the role of discretization in Section \ref{ms-4}.

\section{Transmission}\label{ms-4}

As discussed in Section \ref{ms-2} the polarization of the EM
waves is crucial
for the observation of the LH properties. The electric field $E$ must be
parallel to the wires, and the magnetic field
$H$  must be parallel to the axis of the SRRs. In the numerical
simulations, we treat simultaneously
both polarizations, $\vec{E}\!\parallel\! x$ as well as 
$\vec{E}\!\parallel\! y$. Due to
the non-homogeneity of the
structure, these polarizations are not separated: there is always non-zero
transmission $t_{xy}$ from the $x$ to $y$ polarized wave. As we will
see later, this  effect is
responsible for
some unexpected phenomena. At present, we keep in mind that they must
be included into the
formula for absorption
\be
A_x=1-|t_{xx}|^2-|t_{xy}|^2-|r_{xx}|^2-|r_{xy}|^2
\ee
and in the equivalent relation for $A_y$.

\begin{figure}[t!]
\begin{center}
\epsfig{file=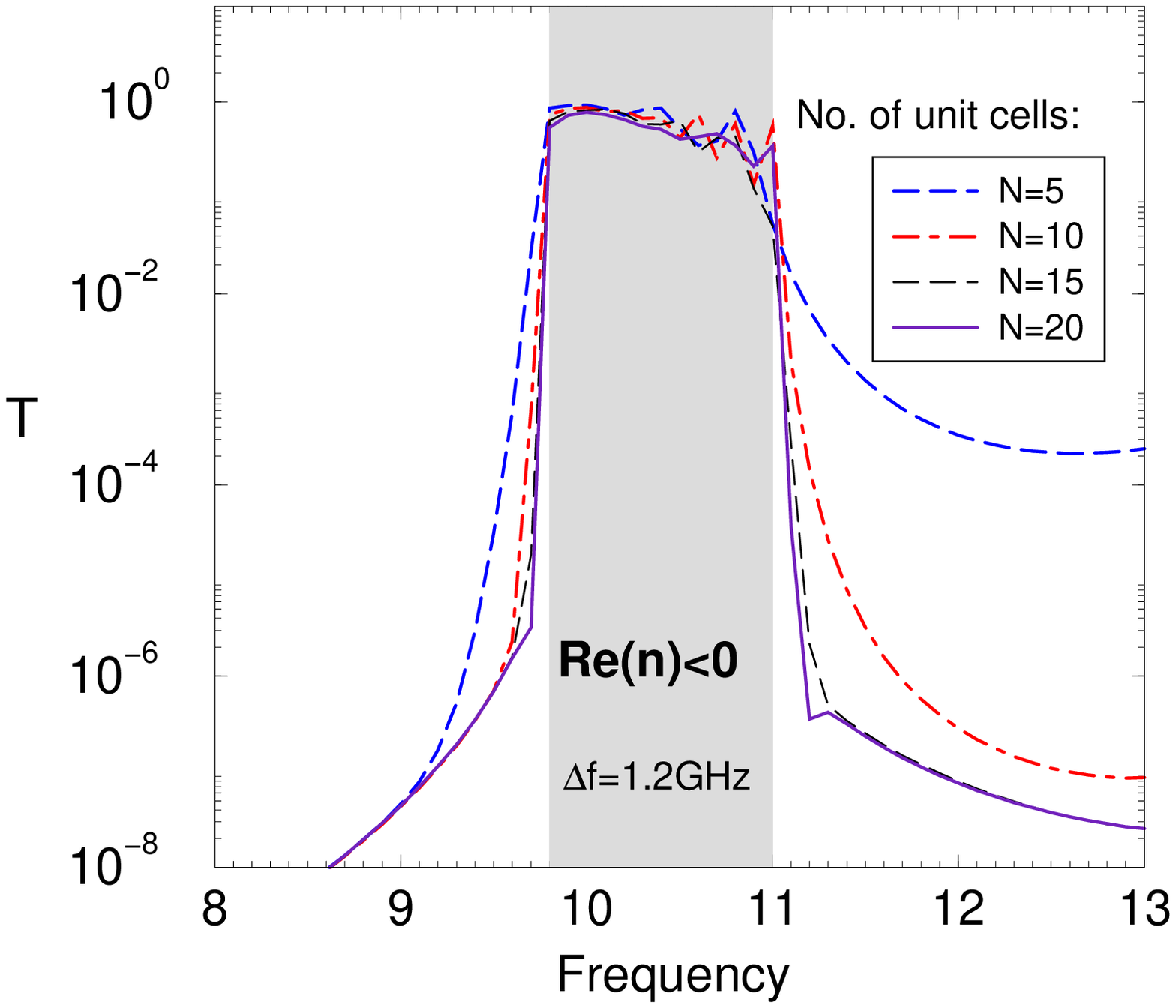,width=0.23\textwidth}
\epsfig{file=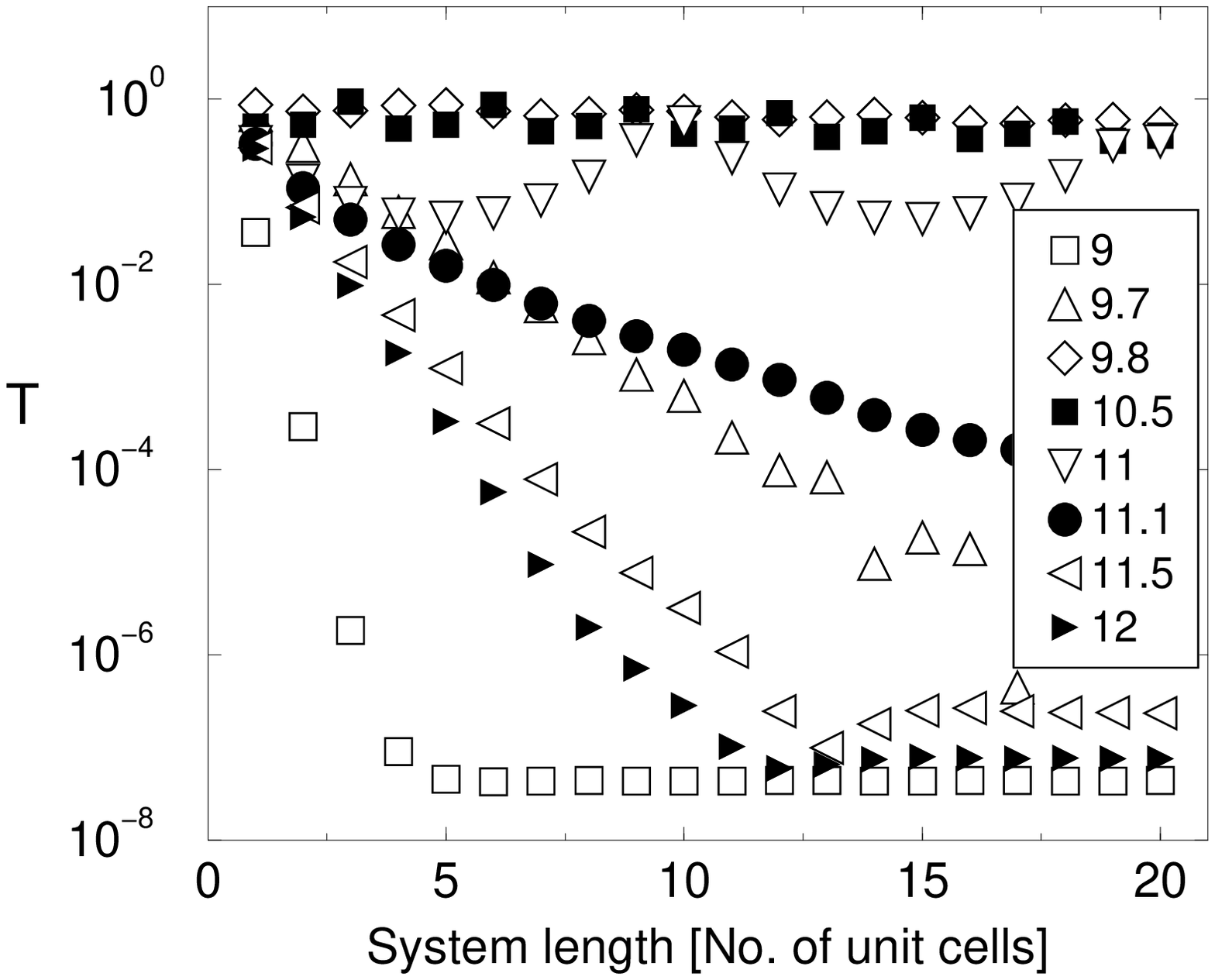,width=0.23\textwidth}
\end{center}
\caption{Transmission power $T$ through the Left-handed meta material
of various lengths. Left: frequency dependence, right: length dependence.}
\label{ms-f5}
\end{figure}

Figure \ref{ms-f5} shows typical data for the transmission in the
resonance frequency region.
A resonance frequency interval, in which the transmission increases
by many orders of magnitude
is clearly visible. Of course, high transmission does not guarantee
negative refraction index.
The sign of $n$ must be obtained by other methods, which will be described in
Sect. \ref{ms-6}.

\begin{figure}[t!]
\begin{center}
\epsfig{file=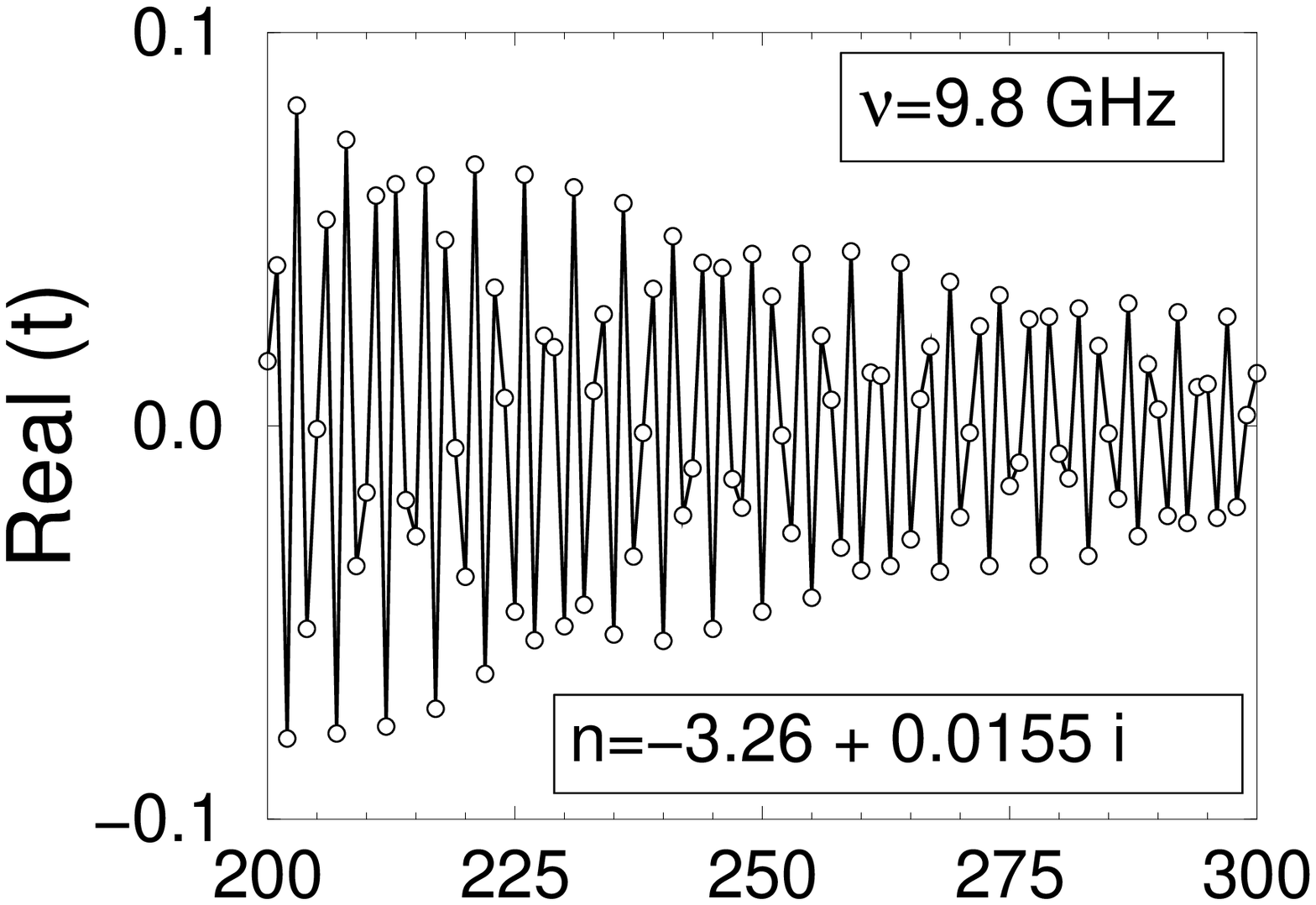,width=0.33\textwidth}\\
\epsfig{file=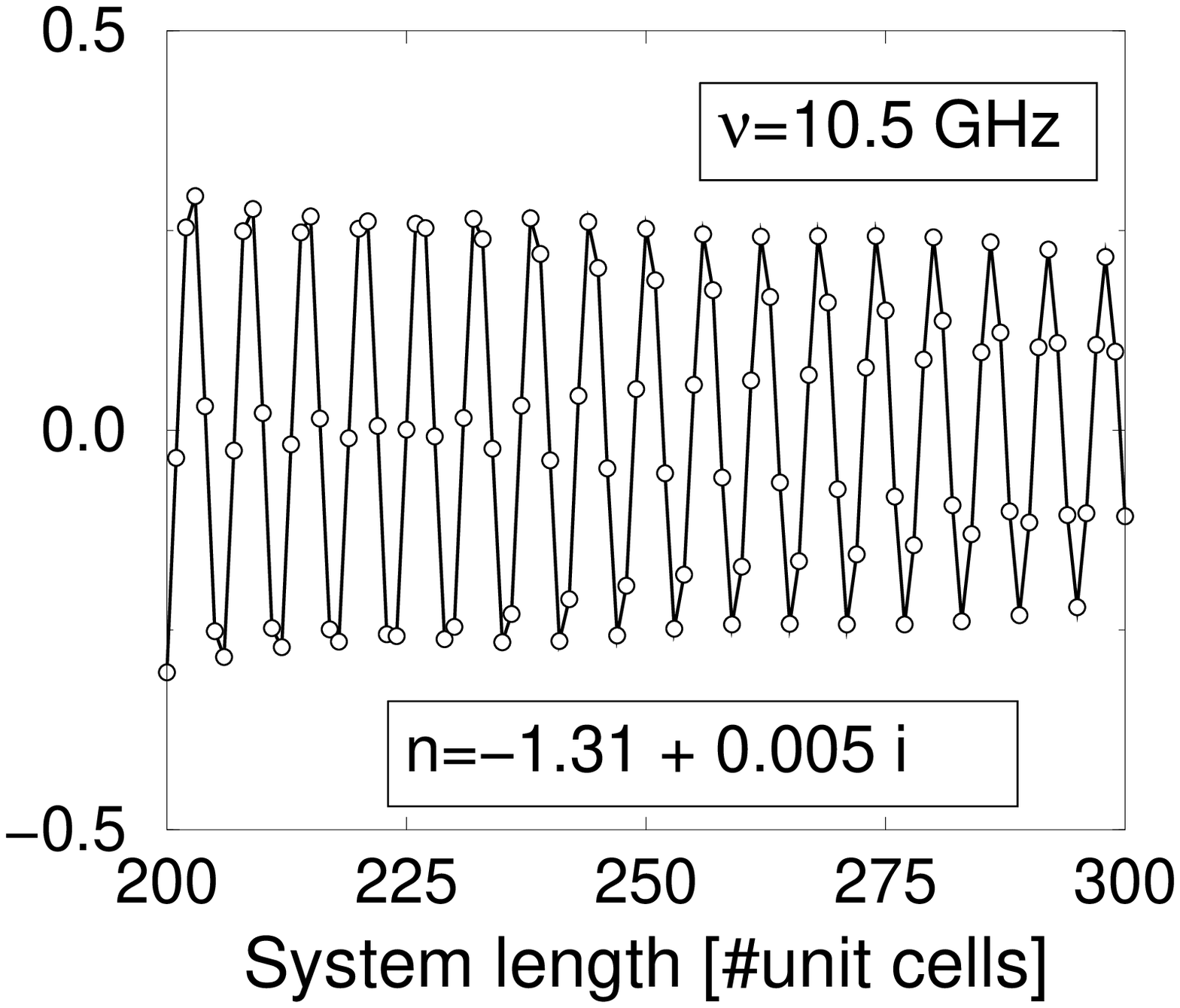,width=0.30\textwidth}\\
\epsfig{file=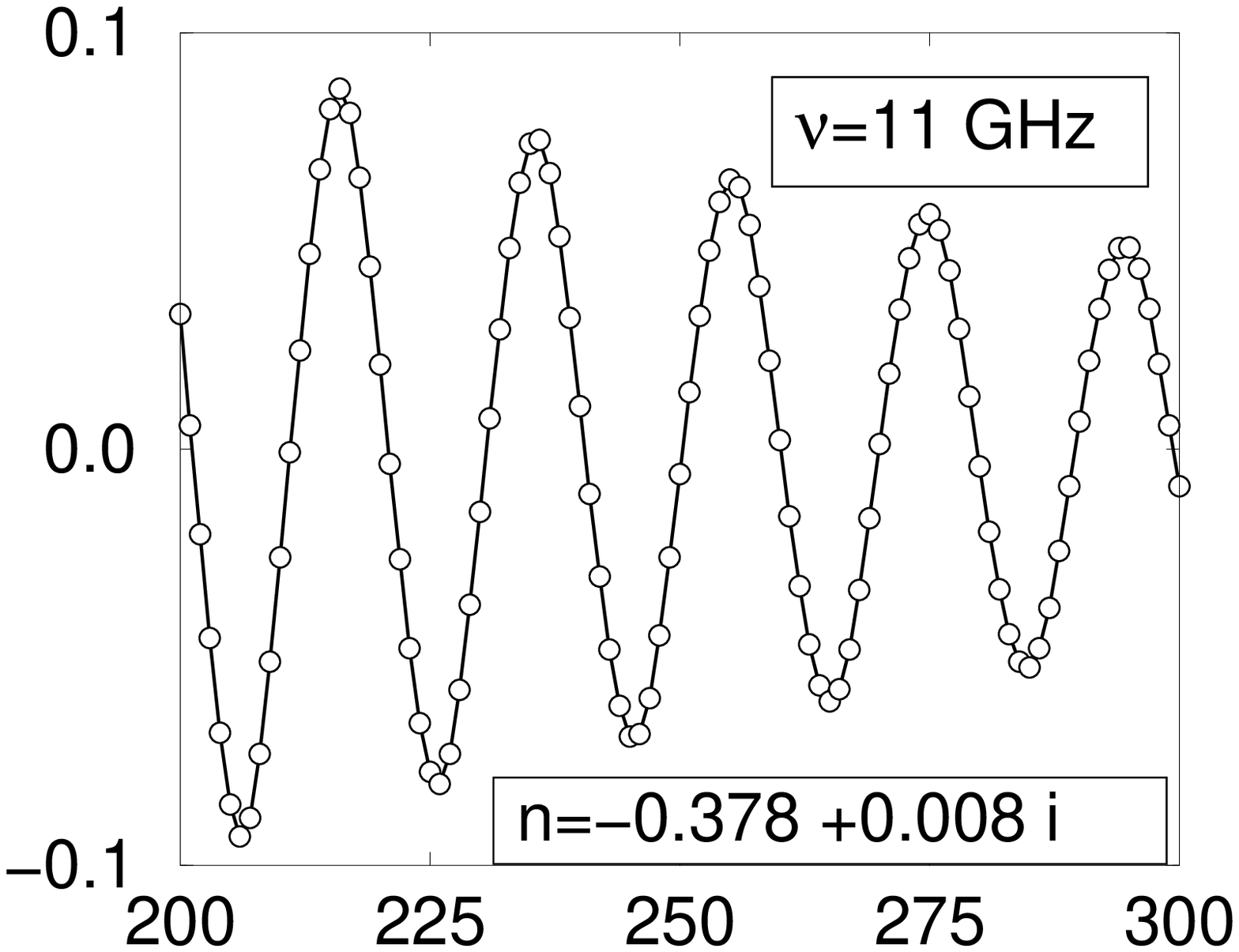,width=0.30\textwidth}
\end{center}
\caption{System length dependence of the real part of the transmission  through
the left-handed structure for
three various frequencies within the left-handed band. The
system length is given as the number of   unit cells
in the propagation direction. Note the different scale on the $y$ 
axis for the three cases.}
\label{ms-f6}
\end{figure}

In contrast to the original experimental data, numerical data show
very high transmission,
indicating that LH structures could be as transparent as the
``classical'' right-handed ones.
This is  surprising, because due to the dispersion, high losses are expected.
Fig. \ref{ms-f6} shows the transmission as the function of the
system length, for three different frequencies inside the resonance 
frequency interval.

Figure \ref{ms-f5} shows also that the transmission never decreases below
a certain limit.
Due to the non-homogeneity  of the structure there is
a non-zero probability $t_{yx}$ that the EM wave, polarized with
$\vec{E}\!\parallel\! y$, is converted into the polarization
$\vec{E}\!\parallel\! x$.
The total transmission $t_{yy}$ consists therefore not only from the
``unperturbed'' contribution
$t^{(0)}_{yy}$, but also from additional terms, which describe the
conversion of the $y$-polarized wave
into $x$-polarized and back to $y$-polarized:
\begin{equation}\label{ttt}
t_{yy}(0,L)=t^{(0)}_{yy}(0,L)+\sum_{z,z'}t_{xy}(0,z)t_{xx}(z,z')t_{yx}(z'L)+\dots~.
\end{equation}
$t^{(0)}_{yy}(0,L)$  decreases   exponentially with the
system length $L$, while  the second term, which represents the
conversion of the $y$-polarized wave
into $x$-polarized wave and back, remains system-length independent, because
$t_{xx}(z,z')\sim 1$ for any distance $|z-z'|$.

\begin{figure}[t!]
\begin{center}
\epsfig{file=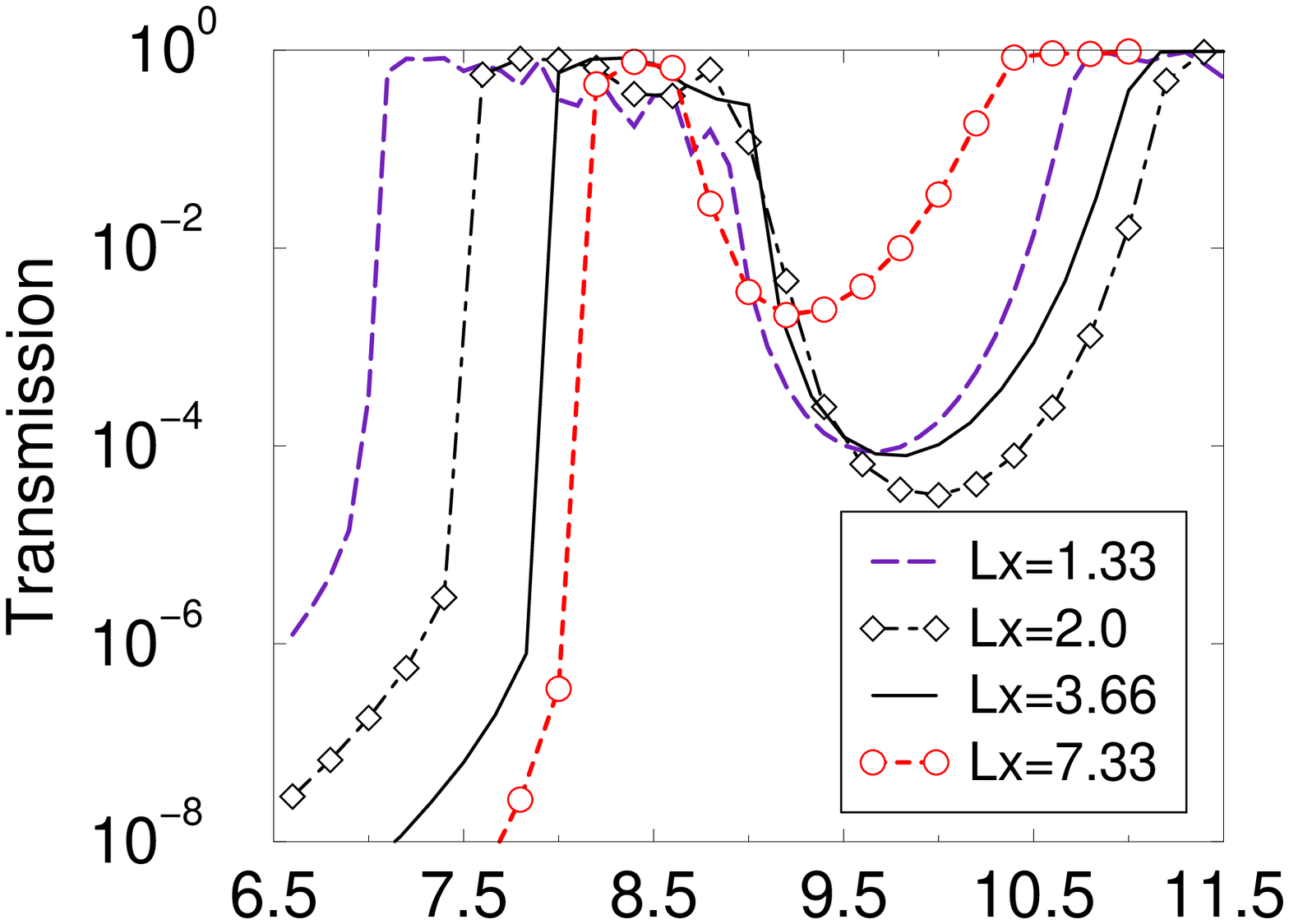,width=0.33\textwidth}\\
\epsfig{file=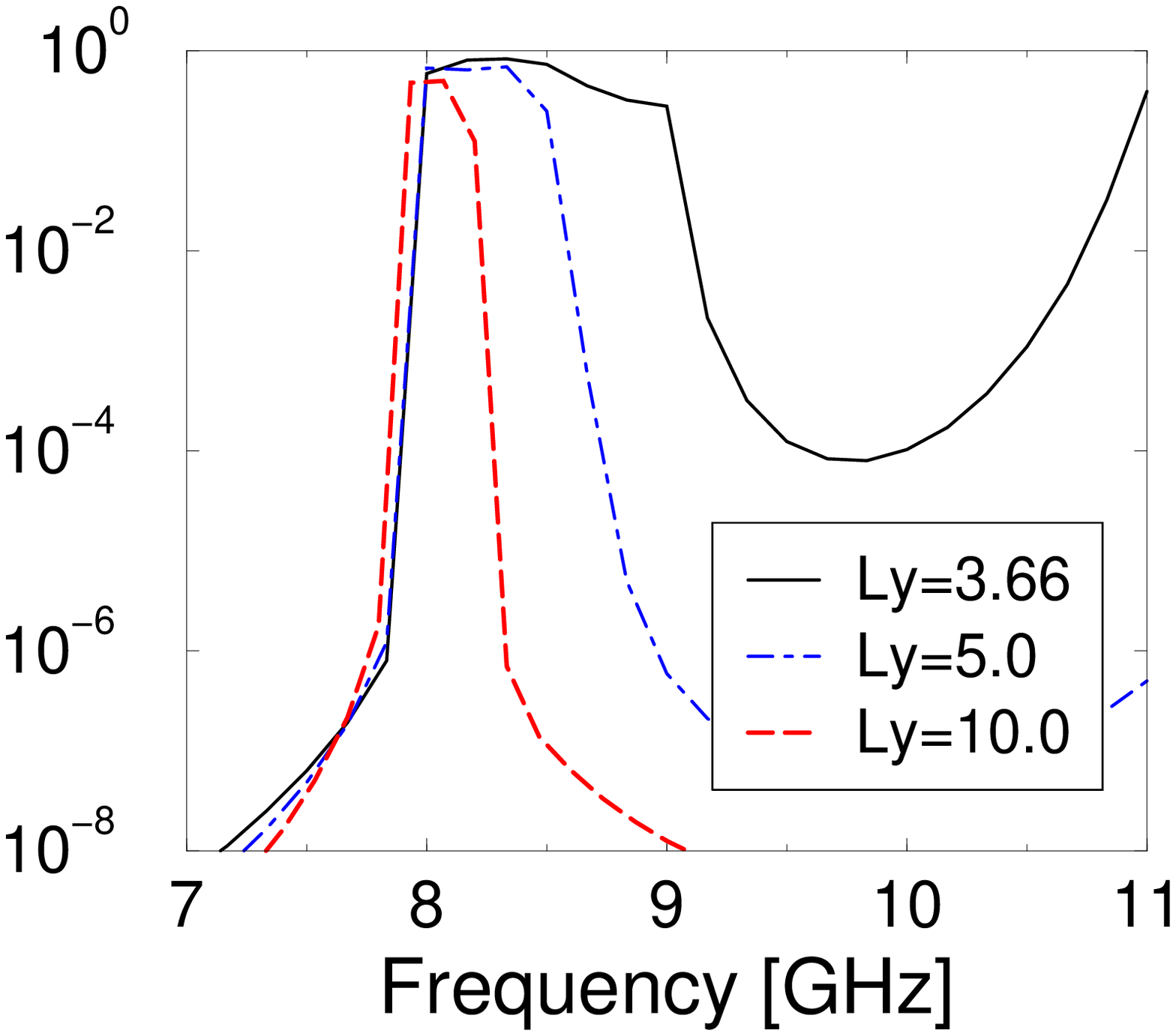,width=0.30\textwidth}\\
\epsfig{file=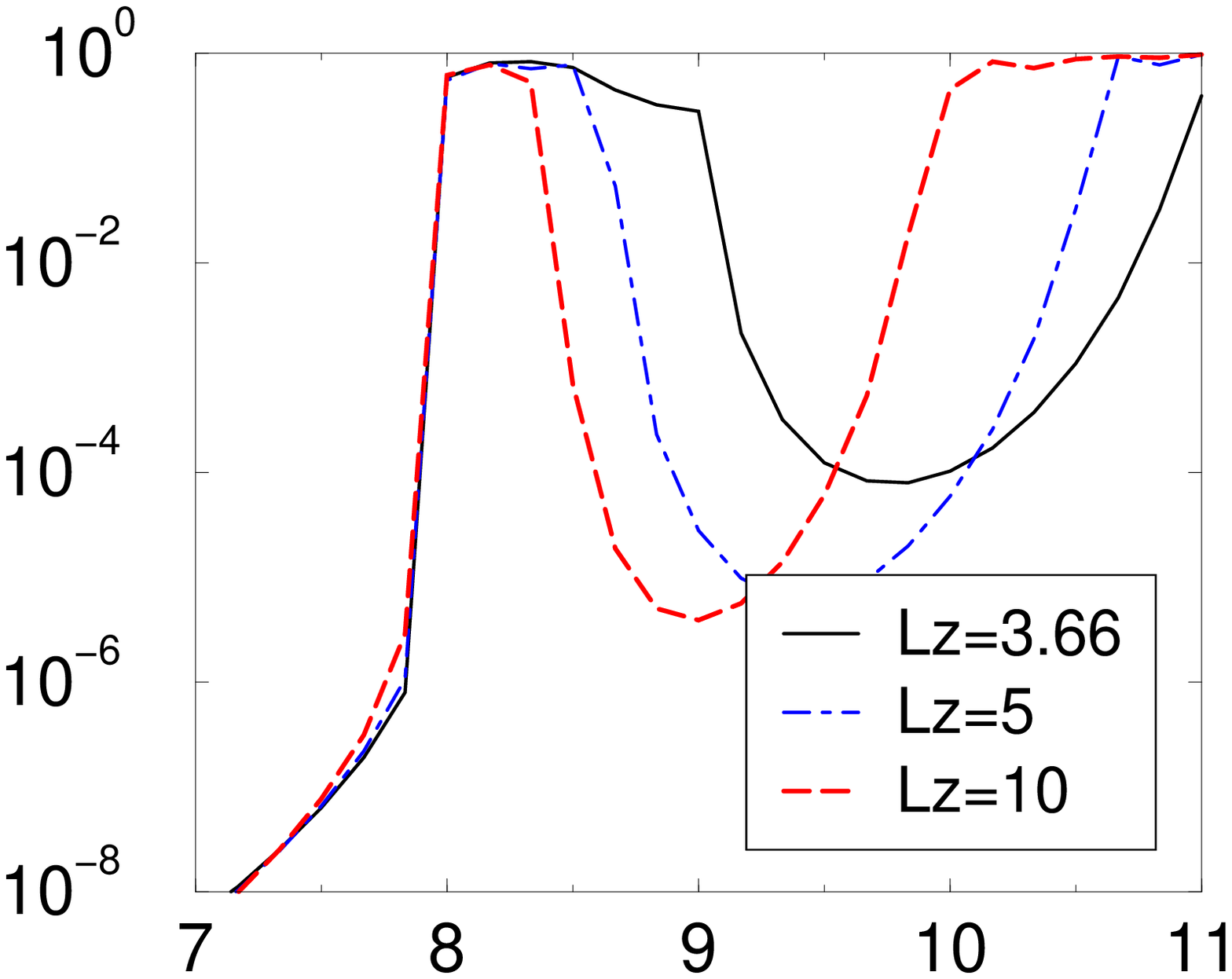,width=0.30\textwidth}
\end{center}
\caption{The transmission peak for various sizes of the unit cell. In 
contrast to the
structure shown in Fig. \ref{ms-f1}, the metallic wire is located on the
opposite side of the dielectric board. This enables to compress the width
of the unit cell as shown in the far left panel.}
\label{ms-f7}
\end{figure}

\section{Structure and transmission}\label{ms-5}

Numerical simulations confirm the existence of the resonance
left-handed frequency
interval. Before we proceed in the calculation of
the effective system parameters,
let us  briefly discuss how the structure of the unit
cell influences
the position and the width of the resonance interval.

\begin{figure}[t!]
\epsfig{file=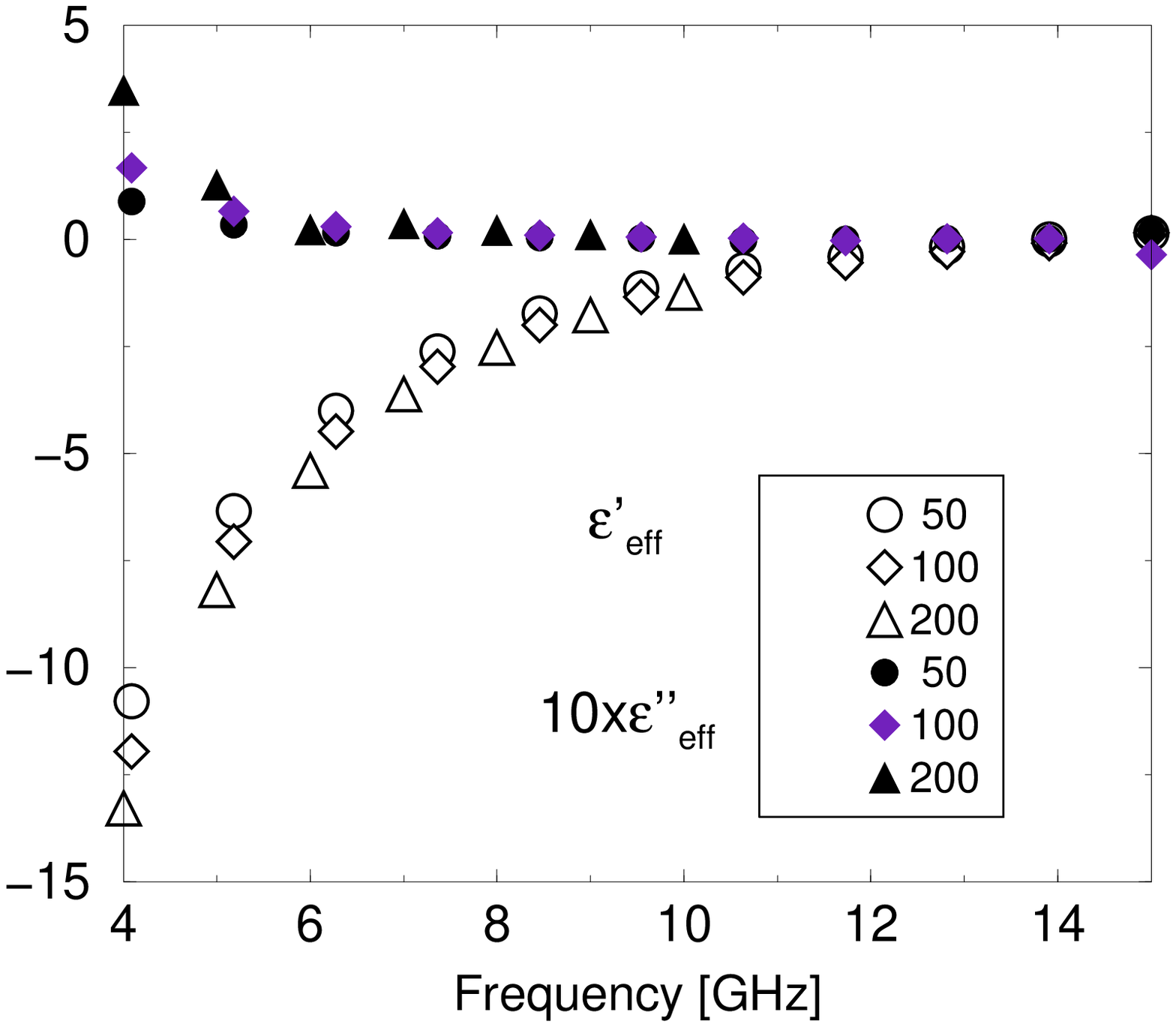,width=0.35\textwidth}
\epsfig{file=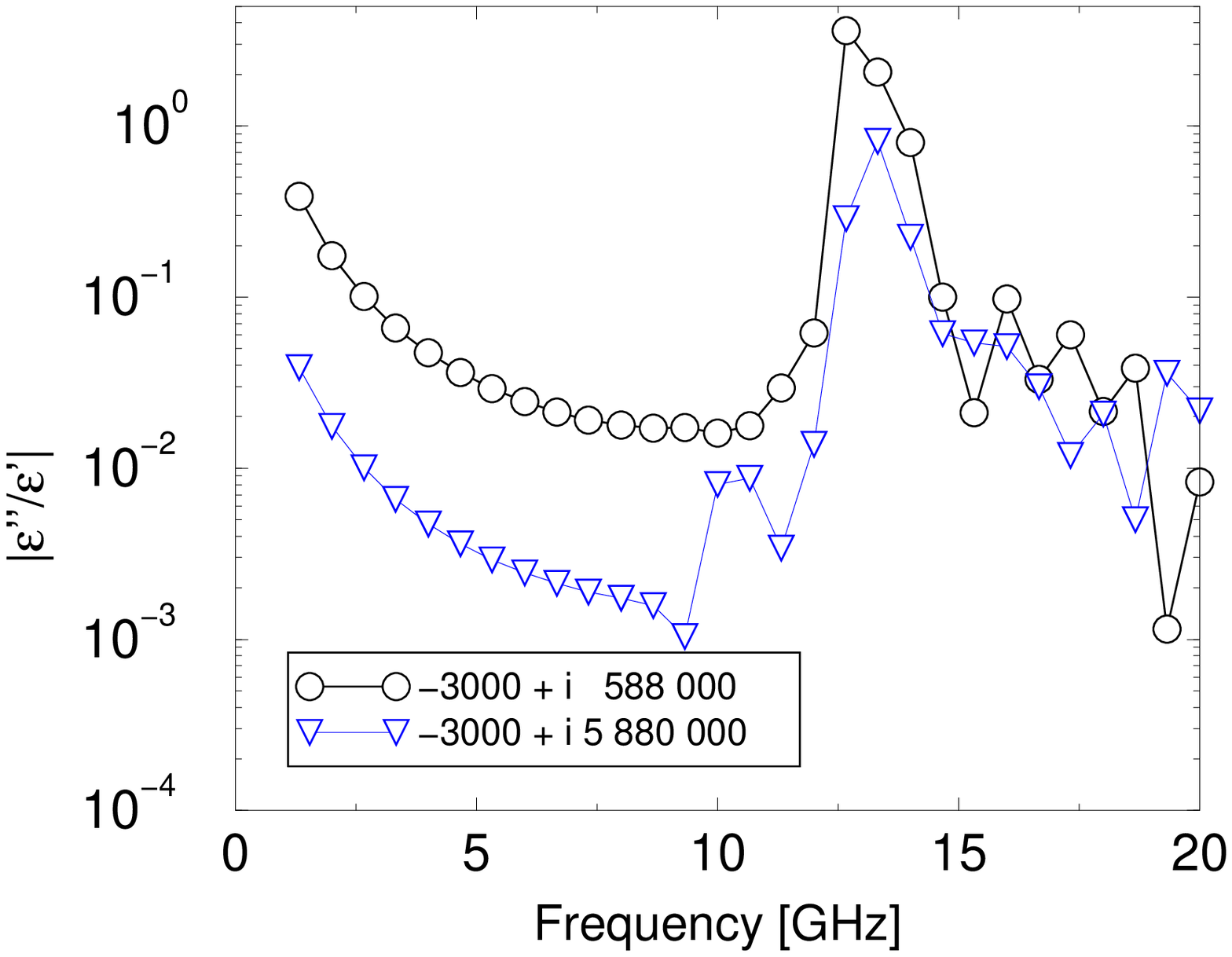,width=0.35\textwidth}
\caption{Top panel:  Effective permittivity $\epse$ for an array
of metallic wires with radius 0.1 mm.  The lattice constant (the 
distance between the wires)
is 5 mm.  The electric field is parallel to the wires.
Note that the imaginary part of the effective permittivity is multiplied
by a factor of 10.
We present results of the analysis of
the transfer matrix data for three different discretization
of the unit cell.
The lower panel shows the ratio $|\eps''/\eps'|$ for an array
of thin metallic wires $100\times 100~\mu$m for two
values of the metallic permittivity. The numerical results prove that 
electromagnetic
losses decrease when the metallic conductance increases.
}
\label{ms-f8}
\end{figure}

The resonance frequency
depends on the structural parameters of the SRR and on the parameters 
of the unit cell
   \cite{Pendry-1}. Qualitative agreement
between the theoretical formulas and the numerical results was
obtained \cite{Weiland,MS-2}.
Another problem is the dependence of the resonance interval on the
size and shape of the
unit cell. As an example, we show in Fig. \ref{ms-f7} how  the
transmission through the
various LH structures depends on the width/high/length of the unit
cell. Notice that the width of the LH frequency interval
increases substantially as shown in the far left panel of fig. \ref{ms-f7}.

Another important parameter is the permittivity $\em$ of the metallic
components. Within the first approximation, we can consider both the 
SRR and the wires
made from a perfect metal. Then both  the conductivity $\sigma$ and  the
imaginary part of the metallic
permittivity $\em''$ are infinite \cite{Jackson}. This option is 
often used in the
simulations of the  commercial
software \cite{Weiland}. In the transfer matrix algorithm, however,
$\em''$ is finite, of the order of $10^5$.
For copper, which is currently used in the experimentally fabricate
LH structures, $\em''\approx 10^7$ \cite{Jackson}.
Some test calculations with higher values of  $\em''$ gave us almost 
the same result,
indicating that $\em''\sim 10^5$ is
already sufficient to simulate realistic materials \cite{MS-4}.

As was mentioned in Sect. \ref{ms-1}, the left-handed structure must 
be dispersive. Dispersion
requires a non-zero imaginary part for the permittivity and
permeability.  Therefore, transmission losses can not be avoided in LH systems.
This seems to be in agreement with the first experimental data:
the transmission measured in the experiments
\cite{Smith,Shelby-APL}, was only of the order $10^{-3}$. Although
recent  experiments reported the transmission very close to unity,
there are still serious doubts in the literature
\cite{Garcia_1,GP} about the possibility to create highly transmitted
LH structures.
Results of numerical simulations, as shown in fig. \ref{ms-f5},
however give that the transmission through the LH structure could be very high,
of order of unity.

The main argument against the expectation of high transmission in LH 
structures  \cite{Garcia_1} is that the
effective permittivity of the array of thin metallic wires is mostly 
imaginary than
real and negative \cite{GP}.
This is a serious objection, because the thickness of the wires in 
fig. \ref{ms-f1} is 1 mm, which exceeds
more than 10 times the realistic parameters of the experimentally 
analyzed structures.
There is no chance to simulate very tiny metallic structures in the 
transfer matrix algorithm.
Fortunately,
Pendry {\it et al.} (formula (36) of \cite{Pendry-JPCM}) had shown 
that the problem of
the thickness of the wires could be
avoided by simultaneously re-scaling  the metallic permittivity.
An increase of the wire radius by factor $\alpha$
could be compensated by  a {\it decrease} of the permittivity of
the metal by factor of $\alpha^2$.  As the metallic
permittivity used in \cite{MS-3} is $\approx 100$ times {\it lower} 
than the realistic permittivity of Cu,
we assume  that the size of the  wires (1 mm) corresponds to Cu wires 
of thickness 0.1 mm, which is close to
size used in experiments ($0.2\times 0.017$ mm).

\begin{figure}[t!]
\begin{center}
\epsfig{file=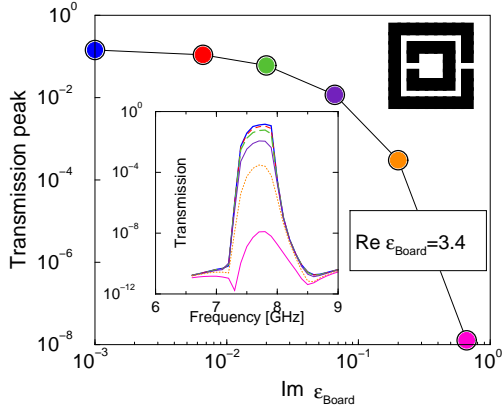,width=0.37\textwidth}
\end{center}
\caption{The dependence of the transmission peak on the imaginary part of
the dielectric permittivity of the dielectric board. Note that
standard dielectrics
have $\eps''\approx 10^{-2}$.}
\label{ms-f9}
\end{figure}

In fig. \ref{ms-f8} we present results for the {\it effective} 
permittivity of an array
of thin metallic wires.
The left panel  confirms  that the transfer matrix algorithm provides us
with realistic data for the transmission and
reflection.  Three different discretizations
of the unit cells were used. For each of them the effective permittivity was
calculated by the method explained in  Sects. \ref{ms-6},
\ref{ms-7}. The right panel of figure \ref{ms-f8} shows that 
transmission losses are
small in realistic LH structures \cite{MS-A}.
Losses in the metallic components of the left-handed structure are 
therefore not responsible for the
low transmission, measured in the experiments  \cite{Smith,Shelby-APL}.
The role of the material properties of the
dielectric board on which SRR is deposited was also studied \cite{MS-4}.
As is shown in Fig. \ref{ms-f9}, very small imaginary part
of $\eps_{\rm Board}$ causes a rapid decrease of the transmission.
This surprising result seems to be in agreement with experiments 
\cite{Claudio-2}. Much higher
transmission was obtained in cases  with extremely
small losses in the dielectric board \cite{Claudio-2}.

\begin{figure}[t!]
\begin{center}
\epsfig{file=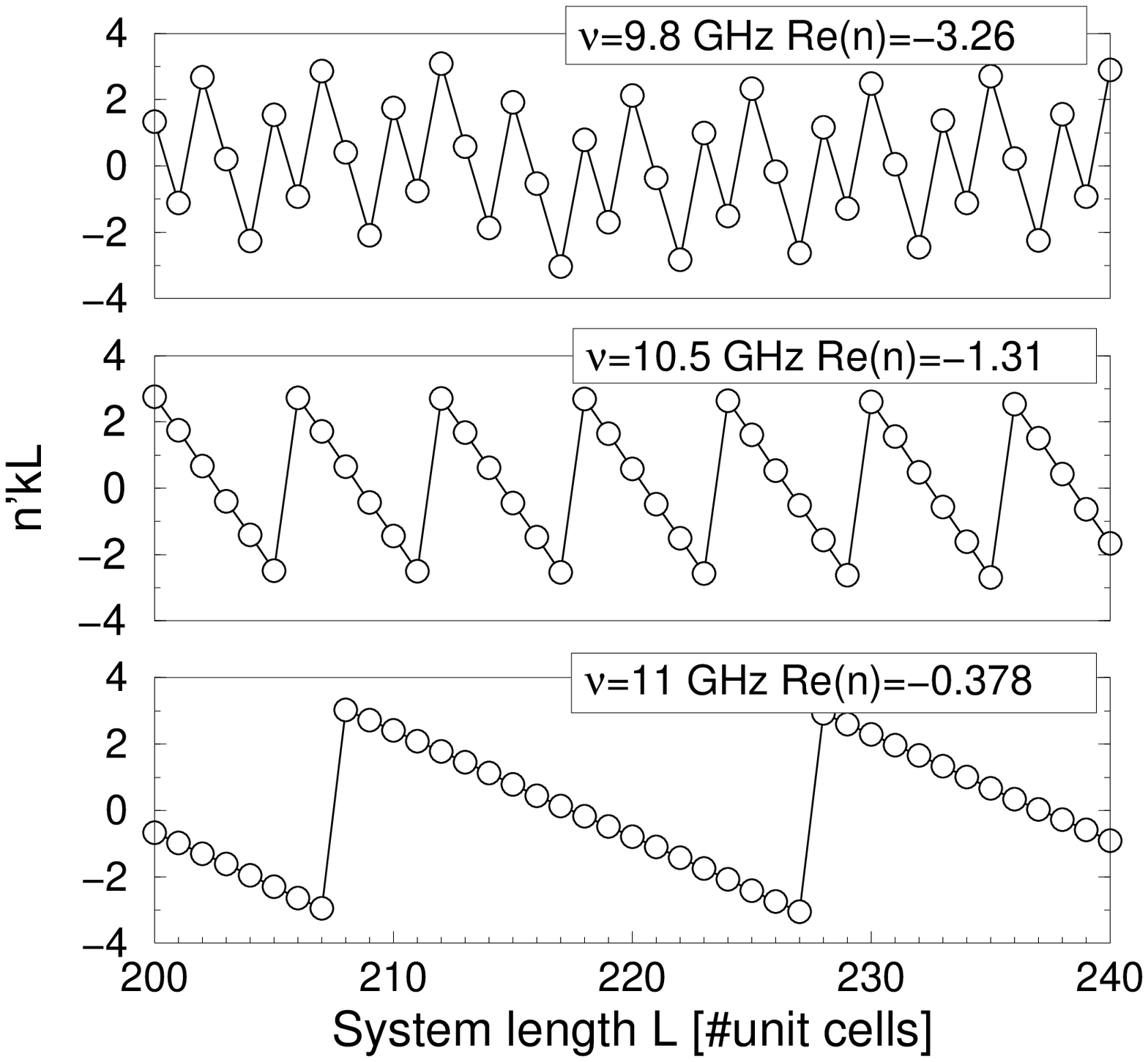,width=0.44\textwidth}
\epsfig{file=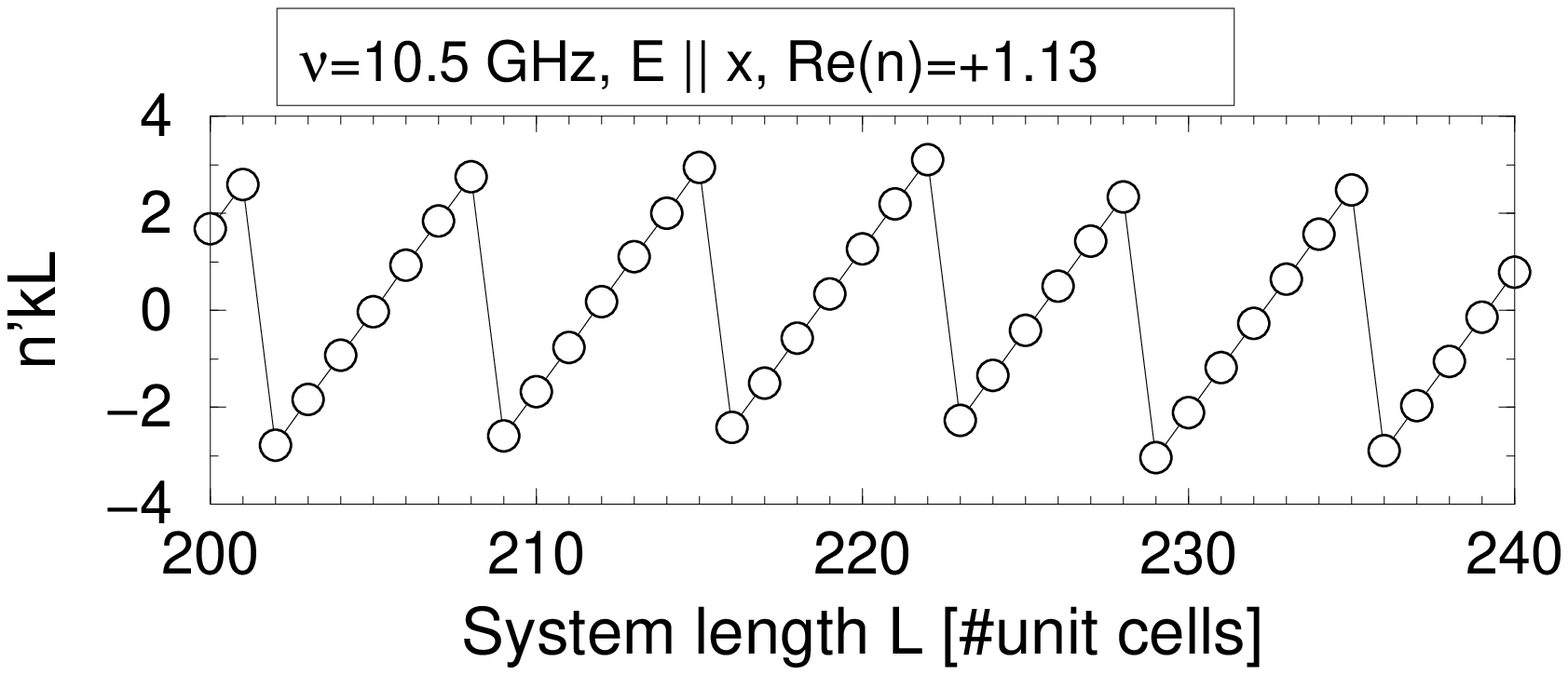,width=0.44\textwidth}
\end{center}
\caption{To panel shows $n'kL$ as a function of the system length $L$ for the EM wave
polarized with $E$ parallel to the wires. The slope
is negative which confirms that $n'$ is negative.
Bottom: 
$n'kL$ as a function of the system length $L$ for the
EM wave polarized $E\parallel x$. The slope
is positive which confirms that Re $n'$ is positive. A 
linear fit gives
that $n'=1.1$. There is almost no interaction of the LH structure with
the EM wave.}
\label{ms-f10}
\end{figure}

\section{Effective index of refraction}\label{ms-6}

As  was discussed above, the structural inhomogeneities of the LH
materials are  approximately ten times
smaller than the wavelength of the EM wave. It is therefore
possible, within a first approximation,
to consider the slab of the LH material as an  homogeneous material.
Then we can  and use the textbook formulas
for the transmission $te^{ikL}$ and reflection $r$ for the homogeneous slab:
\begin{equation}
t^{-1}=\left[
\cos(nkL)-\frac{i}{2}\left(z+\frac{1}{z}\right)\sin(nkL)\right]
\label{tt}
\end{equation}
and
\begin{equation}
\frac{r}{t}=-\frac{i}{2}\left(z-\frac{1}{z}\right)\sin(nkL)
\label{rt}
\end{equation}
Here, $z$ and $n$ are the {\it effective} impedance and the refraction index,
respectively, $k$ is the
wave vector of the  incident EM wave in vacuum, and  $L$ is the length of
the LH slab.
We only consider perpendicular incident  waves, so that only the $z$
components of the effective parameters
are  important.  To simplify the  calculations, we neglect also the
conversion of the
polarized EM wave into another polarization, discussed in the 
Section \ref{ms-4}.
More accurate analysis should  treat both $t$ and $r$ as $2\times 2$ matrices.
Here we assume that the off-diagonal elements of these matrices are negligible:
\be
|t_{xy}|\ll |t_{yy}|
\label{aniz}
\ee

In the present analysis, we use the numerical data for  the 
transmission and the reflection
obtained by the transfer matrix simulation.
The expressions for the transmission and the reflection can be inverted as
\begin{equation}
z=\pm\sqrt{\frac{(1+r)^2-t^2}{(1-r)^2-t^2}}
\label{zz}
\end{equation}
\begin{equation}
\cos(nkL)=X=\frac{1}{2t}\left(1-r^2+t^2\right)
\label{cosn}
\end{equation}
The sign of $z$  is determined by the condition
\be
z'>0~
\label{condz}
\ee
which determines $z$ unambiguously.
The obtained data for $z$ enables us also to check the assumption of
the homogeneity of the system.
We indeed found that $z$ is independent of the length of the system $L$.

The second relation, (\ref{cosn}), is more difficult to invert since
$\cos^{-1}$ is not an unambiguous function. One set of physically
acceptable  solutions is determined
by the requirement
\be
   n''>0~
\label{condn}
\ee
which assures that the  material is passive. The
real part of the refraction index, $n'$, however, suffers from the
unambiguity of $2\pi m/(kL)$ ($m$ is an integer).
To avoid this ambiguity, data for various system length $L$ were  used.
As $n$ characterizes the material property of
the system, it  is  $L$ independent. Using also the
requirement that $n$ should be a continuous function
of the frequency,
the proper solution of (\ref{cosn})
was  found  and the resonance frequency interval, in which $n'$ is 
{\it negative}
was identified \cite{MS-3}.
Here, we use another method for the calculation of $n$ and $z$:
Equation (\ref{cosn}) can be written as
\begin{equation}
e^{-n''kL}\left[\cos(n'kL)+i~\sin(n'kL)\right]=Y=X\pm\sqrt{1-X^2}.
\label{ren}
\end{equation}
Relation (\ref{ren}) enables us to find unambiguously both the real
and the imaginary part
of the refraction index from the linear fit of $n''kL$ and $n'kL$
{\it vs}  the system length $L$.
The requirement (\ref{condn}) determines the sign in the r.h.s. of
Eq. \ref{ren}, because $|Y|<1$.
Then, the linear fit of $n'kL$ {\it vs} $L$ determines unambiguously the
real part of $n'$.

Figure \ref{ms-f10} shows the $L$ dependence of $n'kL$ for three
frequencies inside the resonance interval
(transmission data for these frequencies are presented in fig. \ref{ms-f6}).
The numerical data proves that the real part of the
refraction index,  $n'$, is indeed {\it
negative} in the resonance interval.
For comparison, we present  also $n'kL$ {\it vs} $L$  for the $x$
polarized wave outside the resonance interval.
As expected,  the slope is positive and gives that $n'=1.13$,  which 
is close to unity.

Besides the sign of the real part of the refraction index, the value
of the imaginary part of $n$, $n''$
is important, since it determines the absorption  of the EM waves
inside the sample.
Fortunately, $n''$ is very small, it is only of the order of $10^{-2}$
inside the resonance interval.
As is shown in fig.  \ref{ms-f6}, quite good transmission was
numerically obtained  also  for
samples with  length of 300 unit cells (which corresponds to a length 
of the system 1.1 m).
This result is very encouraging and indicates that  left-handed 
structures could be as transparent as
right-handed materials.

\medskip

There are two constrains in the present method.
(i)
While the above method works very well in the right side of the
resonance interval, we had problems
to estimate $n$ in the neighborhood of the left border of the
resonance region, where
$n'$ is very large and negative. This is, however, not surprising
since in this frequency region
the wavelength of the propagating wave becomes comparable with the
size of the unit cell, so that
the  effective parameters have no physical meaning.
(ii)
Outside the resonance interval, we have serious problems to
recover proper values of $n$. This is due to the conversion of the
$x$ polarized wave into a $y$ polarized.
As a result, we  do not have enough numerical data for obtaining
the $L$-dependence of the transmission $t$. As it is shown in Fig. \ref{ms-f5},
only data for 2-3 unit cells are representative for the 
$\vec{E}\!\parallel\! y$ wave.
The condition (\ref{aniz}) is not any more
fulfilled, and the present theory must be generalized as discussed above.

\begin{figure}[t]
\epsfig{file=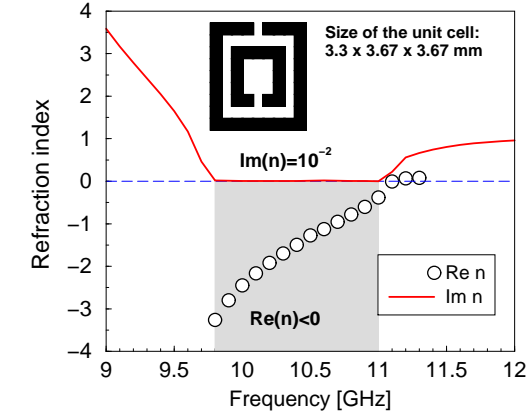,width=0.37\textwidth}\\ \epsfig{file=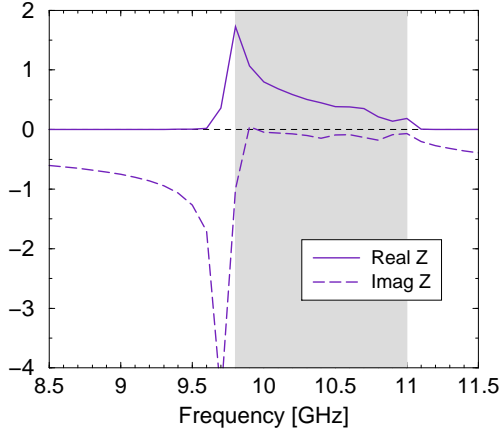,width=0.37\textwidth}\\
\epsfig{file=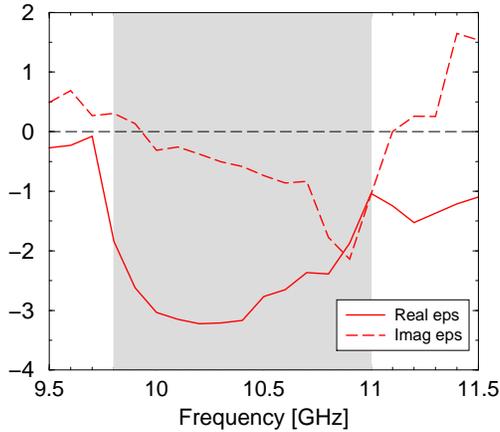,width=0.37\textwidth}\\ \epsfig{file=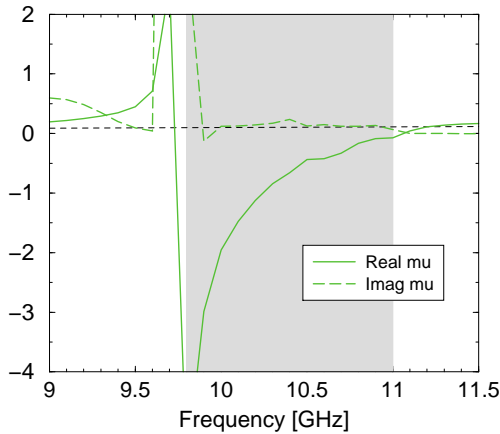,width=0.37\textwidth}
\caption{Refraction index $n$, impedance $z$, permittivity $\eps$ and
permeability $\mu$ as a function of frequency in the resonance
interval.  Close to the resonance frequency $\nu=9.8 GHz$, there are 
problems to
estimate $n$ and $z$, because the wave length of the propagated EM 
wave becomes comparable
with the size of the unit cell  of the left-handed structure.
Dashed area shows the resonance interval.
}
\label{ms-f11}
\end{figure}

\section{Effective permittivity and permeability}\label{ms-7}

Once $n$ and $z$ are obtained
\be
\epseff=n/z\quad\quad{\rm  and }\quad\mueff=nz,
\label{epsmu}
\ee
one  can estimate unambiguously the effective permittivity and
permeability of the investigated
structure. Results are shown in fig. {\ref{ms-f11}.
Frequency dependence of the effective refraction index and
$\epseff'$, $\mueff'$ is also presented
in \cite{MS-3}. Our numerical results confirm the resonance
behavior of the $\mueff'$, in agreement with
theoretical predictions of Pendry {\it et al} \cite{Pendry-1}.
In addition to these
expected results, we also found rather strong electric
response of the SRR, which manifests itself as the  ``anti-resonant''
behavior of $\epseff'$ in the resonance frequency interval
and by the decrease of  the plasma frequency (fig. 4 of ref. 
\cite{MS-3}, see also \cite{OBrien-1}).

The real parts of the permittivity and permeability, $\epseff'$ and 
$\mueff'$ are {\it negative} in the
resonance interval, as expected. Surprisingly, the
imaginary part of the
effective permittivity, $\epseff''$ was also found {\it negative}. This
seems to contradict our physical intuition,
since, following \cite{Landau}, Sect. 80, the electro-magnetic 
losses are given by
\be\label{1}
Q=\frac{1}{2\pi}\int d\omega \omega\left[\eps''|E|^2+\mu''|H|^2\right]
\ee
In passive materials, we require $Q>0$. This is trivially satisfied if
we require that both $\eps''$ and $\mu''$ are positive.
Negative $\epseff''$ was obtained also in other structures \cite{OBrien-1}.

More detailed analysis is also necessary for  the energy of the EM 
wave inside the
left-handed material. The formula for the total energy, given by Eq. \ref{11},
does not assure the positiveness of the energy because
the first term in Eq. (\ref{11}) is negative in the left part of the 
resonance interval.
Indeed, as shown in fig. \ref{ms-f11}, both
$\eps'$ and $\partial\eps'/\partial\omega$ are negative, so that
\be\label{negative}
\frac{\partial(\omega\eps')}{\partial\omega}<0.
\ee

Although formulas (\ref{11},\ref{1}) are not the most general formulas for the
energy of the EM field \cite{Ruppin-2}, we show that  both
$\epseff''<0$ and  relation (\ref{negative}) are consistent with 
(\ref{1}) and (\ref{11}),
respectively.
  We use the definition of the impedance
\cite{Landau}, Sect. 83,
\be\label{2}
E^2=\frac{\mu}{\eps}H^2
\ee
Then, with the help of relations (\ref{epsmu}), we re-write the 
relation (\ref{1}) into the form
\be
Q=\frac{1}{2\pi}\int d\omega \omega |H|^2\times 2n''(\omega)z'(\omega)
\ee
which  assures that $Q$ is positive
thanks to the  conditions that $z'>0$ and $n''>0$ (\ref{condz},\ref{condn}).
There is therefore no physical constrain to the sign of the imaginary 
part of the
permittivity and permeability.

With the help of the relation (\ref{2})
Eq. \ref{11} can be rewritten into the form
\be\label{12}
U=\frac{1}{2\pi}\int d\omega
|H|^2\left[|z|^2\frac{\partial(\omega\eps')}{\partial\omega}
+\frac{\partial(\omega\mu')}{\partial\omega}\right].
\ee
Using the numerical data for the impedance and for real part of the 
permittivity,
we checked that indeed the expression in the bracket of the r.h.s. of 
(\ref{12}) is
always positive.

\section{Further development, unsolved problems and open questions}\label{ms-8}

We reviewed some recent experiments and numerical simulations on the
transmission of the electromagnetic waves through left handed  structures.
For completeness, we note that
recently, Notomi
\cite{Notomi} has studied the light propagation in strongly modulated 
two dimensional
photonic crystals (PC).
In these PC structures the permittivity is periodically modulated in 
space and is {\it positive}.
The permeability is equal to unity.  Such
PC behaves as a material having an effective refractive index 
controllable by the band structure.
For a certain frequency range  it was found by  FDTD simulations
\cite{Notomi,Voula-1,Voula-2} that $n_{\rm eff}$ is negative.
  It is important to examine if left-handed behavior can be observed in photonic
crystals at optical frequencies.

Negative refraction on the interface of a three dimensional PC 
structure has been observed experimentally
by Kosaka {\it et al.} \cite{Kosaka} and a negative refractive index 
associated to that was reported.
  Large beam steaming has been observed in \cite{Kosaka}, that authors 
called ``the superprism phenomena''.
  Similar unusual light propagation has been observed in 
one-dimensional and two dimensional
refraction gratings. Finally, a theoretical work
\cite{Gralak} has predicted a negative refraction index in photonic crystals.

\medskip

Studies of the left handed structures open a series of new 
challenging problems for theoreticians
as well as for experimentalists.
The  complete  understanding of the properties of left-handed 
structures requires
the reevaluation of
some ``well known'' facts of  the electromagnetic theory.  There is no
formulas with {\it negative} permeability in  classical
textbooks of electromagnetism \cite{Landau,Jackson}. Application of 
the existing formulas
to the analysis of
left handed  structures  may lead to some strange
results. The theory of EM field has to be reexamined assuming 
negative $\mu$ and $\epsilon$.
We need to understand completely the relationship
between the real and the imaginary parts of the permittivity and the 
permeability.
Kramers-Kronig relations
should be valid, but nobody have verified them yet in the case of the 
left-handed structures.
The main problem is that we need $\epseff$
and $\mueff$ in the entire range of frequencies, which is difficult 
to obtain numerically.
Then,  due to the anisotropy of the structure as well as the nonzero 
transmission $t_{xy}$ in Eq. (\ref{tt}),
Kramers-Kronig relations should be generalized. We do not believe that today's
numerical data enables their verifications with sufficient accuracy.

Both in the photonic crystals and LHM literature there is a lot of 
confusion about
what is the correct definitions of the phase and group refractive 
index and what is their
  relations to negative refraction.
In additions, it is instructive to see how the LH behavior is related 
with the sign of the
phase and group refractive indices for the PC system.
The conditions of obtaining LH behavior in PC were recently examined 
in \cite{Voula-1}.
It was demonstrated that the existence of negative refraction is 
neither a prerequisite
nor guarantees a negative effective refraction index and so LH 
behavior. Contrary,
{\it  LH behavior can be seen only if phase refractive index $n_{\rm 
phase}$ is  negative}.
Once $n_{\rm phase}$ is negative, the product $\vec{S}.\vec{k}$ is 
also negative, and the vectors
$\vec{k}$, $\vec{E}$ and $\vec{H}$ form a left handed set, as 
discussed in the Introduction.

Problems of causality arises also in connection with ``well known'' 
and accepted relations like
$\partial (n'(\omega)\omega)/\partial\omega>0$. It is evident, that 
this relation
can not be valid in the vicinity
of the left border of left-handed frequency interval, since both $n'$ 
and $\partial n'/\partial\omega $ are
negative. The same problem arises for the real part of $\epseff$. Also relation
(\ref{negative})  requires more exact and complete treatment.
We need more general relations for the energy
of the EM field \cite{Ruppin-2}  which incorporates all the allowed 
signs of the real and imaginary
parts of the permittivity and permeability.

Following  problems that are currently discussed in literature.
The negative Snell's law requires
the  understanding in more detail of the relationship between the 
Poynting vector,
the group velocity, and the phase velocity.
We believe, that there is no controversial in this phenomena
\cite{Pendry-answer}.
Anisotropy of real left-handed materials inspires further development 
of super-focusing
\cite{SS}.

We believe that further analysis, of what happens when EM wave 
crosses the boundary
of the left-handed and right-handed systems, will bring more
understanding  of the negative refraction
as well as perfect lensing.
Numerical FDTD simulations of the transmission of the  EM wave through
the interface of the positive and negative refraction index \cite{Voula-2}
showed that the wave is trapped temporarily
at the interface and after a long time the wave front moves
eventually in the direction of negative refraction.
Computer simulations of the transmission through LH wedge 
\cite{Claudio-private}
also confirm   that EM wave spends some time on the boundary before 
the formation of the left-handed
wave front.
Formation of surface waves
\cite{Ruppin,Haldane,Gomez}
can explain ``perfect lensing''  without violating causality.
Recent development, both numerical \cite{Kik} and experimental
\cite{Cubukcu} indicate that perfect (although not absolutely 
perfect) lensing might
be possible.
We need also to understand some peculiar properties of the left 
handed structures
due to its anisotropy \cite{Hu} and bi-anisotropy \cite{Marques}.

Last but not least,
let us mention an attempt to find new left-handed structures, both in 
the traditional
metallic left-handed  structures \cite{Gorkunov}
and in the photonic crystals  \cite{OBrien-1,OBrien-2}.

\medskip
\noindent{\bf Acknowledgments} We want to thank D.R. Smith,
E.N. Economou, S. Foteinopoulou and I. Rousochatzakis for fruitful 
discussions. This work was supported
by Ames Laboratory (contract. n. W-7405-Eng-82). Financial support of DARPA,
NATO (Grant No. PST.CLG.978088) and APVT (Project n. APVT-51-021602) 
and EU project DALHM
is also acknowledged.


\begin{thebibliography}{99}
\bibitem{pozn} {Besides the name ``Left handed materials'' (metamaterials), 
this new structures are  also called  ``Negative index  materials'' 
(NIM),
``Double negative metamaterials'' (DNM) or ``Backward-wave medium'' 
in the literature.}
\end{thebibliography}
\end{document}